\documentstyle[12pt]{article}

\begin{document}

\begin{titlepage}

\begin{center}

{\Large \bf
Weyl-Wigner-Moyal Formulation of a Dirac Quantized Constrained System}
\vskip .6in

Domingo J. Louis-Martinez
\vskip .2in

Science One Program and\\ 
Department of Physics and Astronomy,\\ University of British
Columbia\\464-6356 Agricultural Road\\Vancouver, Canada

\end{center}
 
\vskip 3cm
 
\begin{abstract}
An extension of the Weyl-Wigner-Moyal formulation of quantum 
mechanics suitable for a Dirac quantized constrained system 
is proposed. In this formulation, quantum observables are
described by equivalent classes of Weyl symbols. The Weyl
product of these equivalent classes is defined. The new
Moyal bracket is shown to be compatible with the Dirac
bracket for constrained systems.
\end{abstract} 
\end{titlepage}

The Weyl-Wigner-Moyal (WWM) method \cite{weyl}-\cite{moyal} provides an
alternative way of formulating quantum mechanics. In this formulation,
quantum observables are described by real functions in phase space. For a
modern introduction to the WWM approach see \cite{osborn,little}.

In \cite{muller} Muller introduced a new product rule for gauge invariant
Weyl symbols and gave a generalizalion of the Moyal formula to the case of
non-vanishing electromagnetic fields. A WWM formulation for systems with
first-class quantum constraints was given in \cite{antonsen}. In
Antonsen's formulation, second-class constraints are turned into
first-class quantum constraints by introducing negative powers of Planck's
constant in the expansions of some Weyl symbols.

Our objective in this paper is to present a WWM formulation of a Dirac
quantized constrained system. We assume that our system may have first-
and second-class constraints. We also assume that gauge conditions have
been chosen, and we treat them as ordinary constraints in phase space.
A classification of second-class constraints within the total hamiltonian
approach can be found in \cite{cabo,lusanna}

We consider a physical system described by the canonical hamiltonian
$H_{c}(z)$ and the
constraints (including the gauge conditions) $\Phi_{\mu}(z)$ ($\mu =
1,2,...,2m$). The phase space of the system is assumed to be the
symplectic
euclidean vector space $\Re^{2n}$. Denote by $z \equiv
(q_{1},p_{1},q_{2},p_{2}, ... ,q_{n},p_{n})$ the points belonging to
$\Re^{2n}$.  The constraints $\Phi_{\mu}(z) = 0$ ($\mu = 1,2,...,2m$)
($m<n$) define a subspace of $\Re^{2n}$ called \cite{gitman} the reduced
phase space $\cal{M}$. 

We assume the constraints satisfy the following conditions:

\begin{equation}
rank\|\frac{\partial \Phi_{\mu}}{\partial z_{i}}\| \stackrel{\cal{M}}{=}
2m
\label{eq1}
\end{equation}

\begin{equation}
det\| \left \{ \Phi_{\mu},\Phi_{\nu} \right \} \| \neq 0
\label{eq2}
\end{equation}

The label $\cal{M}$ above the equality indicates that this equality holds
in the subspace $\cal{M}$.
Condition (\ref{eq1}) guarantees that the dimension of the reduced phase
space $\cal{M}$ is $2n-2m$. Condition (\ref{eq2}) tells us that
$\Phi_{\mu}$
($\mu = 1,2,...,2m$) form a set of second-class constraints \cite{dirac}.
The bracket in (\ref{eq2}) is a Poisson bracket in $\Re^{2n}$. The
Poisson bracket between any two differentiable functions $f$ and $g$ in
$\Re^{2n}$ can be written as \cite{osborn}:

\begin{equation}
\{f,g\} = \sum_{i,j=1}^{2n} J_{ij}\nabla_{i} f \nabla_{j} g,
\label{eq3}
\end{equation}

\noindent where,

\begin{equation}
J = \left( \begin{array}{ccccccc} 0 & 1 &  & &  & & 
\\-1 & 0 &   &  &  &  &  \\ & & 0 & 1 &  &  & 
\\ &  & -1 & 0 &  &  & 
\\ &  &  &  & ... &  &
\\ &  &  &  & & 0 & 1
\\ &  &  &  &  & -1 & 0
\end{array} \right)
\label{eq4}
\end{equation}

\vskip 1cm

\noindent is the symplectic matrix in $\Re^{2n}$ \cite{osborn}.

For systems with second-class constraints, it is useful to introduce the
Dirac bracket \cite{dirac}. The Dirac bracket between two differentiable
functions $f$ and $g$ on $\Re^{2n}$ is a function defined as follows
\cite{dirac}:

\begin{equation}
\{f,g\}_{D} = \{f,g\} - \sum_{\mu,\nu =1}^{2m}\{f, \Phi_{\mu}\}
C^{-1}_{\mu\nu}\{\Phi_{\nu}, g\}
\label{eq5}
\end{equation}

\noindent where,

\begin{equation}
C_{\mu\nu} \equiv \{\Phi_{\mu},\Phi_{\nu}\}
\label{eq6}
\end{equation}

A discussion of the properties and applications of Dirac brackets can be
found in \cite{dirac}. Dirac brackets satisfy the Jacobi identity
\cite{dirac}:

\begin{equation}
\{\{f,g\}_{D},h\}_{D} + \{\{h,f\}_{D},g\}_{D} + \{\{g,h\}_{D},f\}_{D} = 0
\label{eq7}
\end{equation}

It is convenient to rewrite the Dirac bracket (\ref{eq5}) in the form:

\begin{equation}
\{f,g\}_{D} = \sum_{i,j=1}^{2n} J^{D}_{ij} \nabla_{i} f \nabla_{j} g
\equiv \nabla f J^{D} \nabla g
\label{eq8}
\end{equation}

\noindent where,

\begin{equation}
J^{D}_{ij} \equiv \{z_{i}, z_{j}\}_{D}
\label{eq9}
\end{equation}

The matrix $J^{D}$ is a degenerate matrix of rank $2n - 2m$. Notice that
the $2m$ vectors 
$\frac{\partial \Phi_{\mu}}{\partial z}$
are null eigenvectors of $J^{D}$. 

\begin{equation}
J^{D}_{ij} .  \frac{\partial \Phi_{\mu}}{\partial z_{j}} = 0,
\hspace{2.0cm}\
\mu = 1,2,...,2m.
\label{eq10}
\end{equation}

Under canonical transformations in $\Re^{2n}$, the rank of the matrix
$J^{D}$ remains invariant.

The evolution of the physical system is described by the 
hamiltonian:

\begin{equation}
H = H_{c}(z) + \sum_{\mu =1}^{2m} \lambda_{\mu}(t) \Phi_{\mu}(z)
\label{eq11}
\end{equation}

The Dirac-Hamilton equations can be written as:

\begin{equation}
\dot{z}_{i} = \sum_{j=1}^{2n} J_{ij} \nabla_{j} H_{c} + \sum_{j=1}^{2n}
J_{ij} \sum_{\mu =1}^{2m} \lambda_{\mu}(t) \nabla_{j} \Phi_{\mu}
\label{eq12a}
\end{equation}

\begin{equation}
\Phi_{\mu}(z) = 0
\label{eq12b}
\end{equation}

The Lagrange multipliers $\lambda_{\mu}$ can be expressed as functions of
the canonical variables in the form \cite{dirac}:

\begin{equation}
\lambda_{\mu} = - \sum_{\nu =1}^{2m} C^{-1}_{\mu \nu} \{\Phi_{\nu},
H_{c}\}
\label{eq13}
\end{equation}

Therefore, the Dirac-Hamilton equations (\ref{eq12a},\ref{eq12b}) can be
written
in terms
of Dirac brackets as follows \cite{dirac}:

\begin{equation}
\dot{z}_{i} = \{z_{i}, H_{c}\}_{D},
\label{eq14a}
\end{equation}

\begin{equation}
\Phi_{\mu}(z) = 0
\label{eq14b}
\end{equation}

The Dirac quantization for systems with second-class constraints consists
in substituting the canonical variables $z$ by quantum operators
$\hat{z}$
that satisfy the quantum constraints:

\begin{equation}
\hat{\Phi}_{\mu} = 0
\label{eq15}
\end{equation}

\noindent and have the commutation rules:

\begin{equation}
[ \hat{z}_{i}, \hat{z}_{j}] = i \hbar \hat{J}^{D}_{ij}
\label{eq16}
\end{equation}

The quantum evolution of the system, in the Heisenberg picture, is
described by the quantum hamiltonian $\hat{H}_{c}$:

\begin{equation}
\frac{d \hat{z}_{i}}{d t} = - \frac{i}{\hbar} [\hat{z}_{i}, \hat{H}_{c}]
\label{eq17}
\end{equation}

The Dirac quantization prescription is applicable to many systems of
interest in Physics \cite{gitman}.

Our purpose is to develop a WWM formulation of a Dirac quantized
constrained system. For simplicity, let us assume that the constraints are
linear functions of $z$:

\begin{equation}
\Phi_{\mu} = \sum_{i=1}^{2n} \alpha_{\mu i} z_{i}
\label{eq18}
\end{equation}

\noindent where, $\alpha_{\mu i}$ are constants such that
(\ref{eq1},\ref{eq2}) are
satisfied.

This is a very strong assumption which is equivalent to assuming that the
reduced phase space $\cal{M}$ is flat: $\cal{M}$ $=$ $\Re^{2n-2m}$.

The first step towards a WWM formulation in $\Re^{2n}$ is to define
physical observables as Weyl symbols. Weyls symbols are real functions on
$\Re^{2n}$. Let us say that two Weyl symbols $A_{1}$ and $A_{2}$ belong to
the same equivalent class ${\cal A}$ if and only if:

\begin{equation}
A_{1} \stackrel{\cal{M}}{=} A_{2}
\label{eq19}
\end{equation}

In our formalism, a quantum physical observable $\hat{A}$ will be
described by a whole equivalent class of Weyl symbols $\cal{A}$. In other
words, in this formulation, the values taken by a Weyl symbol outside the
reduced phase space $\cal{M}$ are of no physical significance.

We need to establish a connection between operators and equivalent classes
of Weyl symbols that describe one and the same physical observable. We
propose the following correspondence:

\begin{equation}
\hat{A} = (2 \pi)^{-2n} \int d\mu(u) d\mu(\zeta) A(\zeta) e^{-iu\zeta}
e^{iu\hat{z}}
\label{eq20}
\end{equation}

\noindent where the measure,

\begin{equation}
d\mu(z) = dz  (2 \pi)^{m} \prod_{\gamma =1}^{2m} \delta(\Phi_{\gamma}(z))
det C^{1 \over 2}
\label{eq21}
\end{equation}

$C$ is the matrix defined in (\ref{eq6}). $\delta$ is the Dirac delta
function.

The operators $\hat{z}_{i}$ satisfy the commutation rules:

\begin{equation}
[ \hat{z}_{i}, \hat{z}_{j}] = i \hbar J^{D}_{ij}
\label{eq22}
\end{equation}

We say that the Weyl symbol $A(\zeta)$ and the operator $\hat{A}$ in 
(\ref{eq20}) describe the same quantum observable. Notice that from
(\ref{eq20},\ref{eq21})
it follows that all Weyl symbols belonging to one equivalent class 
$\cal{A}$ correspond to the same quantum observable $\hat{A}$.

Notice also that:

\begin{equation}
(2 \pi)^{-2n} \int d\mu(u) d\mu(\zeta) \zeta_{i} e^{-iu\zeta}
e^{iu\hat{z}} = \hat{z}_{i}
\label{eqA}
\end{equation}

From (\ref{eqA}), it immediately follows that:

\begin{equation}
\sum_{\mu=1}^{2m} \alpha_{\mu i} \hat{z}_{i} = 
(2 \pi)^{-2n} \int d\mu(u) d\mu(\zeta)
\sum_{\mu=1}^{2m}\alpha_{\mu i} \zeta_{i} e^{-iu\zeta}
e^{iu\hat{z}} = 0
\label{eqB}
\end{equation}

Let us define the Weyl product of two equivalent classes $\cal{A}$ and
$\cal{B}$. This product is also an equivalent class of symbols. Let $A$
and $B$ be two Weyl symbols. The Weyl symbol $A * B$ will be given by the
formula:

\begin{equation}
(A*B)(z) \stackrel{\cal{M}}{=} (\pi \hbar)^{-2n} ({\hbar \over 2})^{2m}
\int d\mu(\zeta) d\mu(\xi) A(\zeta) B(\xi) e^{{2i \over \hbar} (\zeta - z)
J^{D} (\xi - z)}
\label{eq23}
\end{equation}

It is not difficult to prove that:

\begin{equation}
\hat{A} \hat{B} = (2 \pi)^{-2n} \int d\mu(u) d\mu(\zeta) (A*B)(\zeta)
e^{-iu\zeta}
e^{iu\hat{z}}
\label{eqC}
\end{equation}

Our definition of the Weyl multiplication of symbols is independent of
what particular representatives are used. Therefore, (\ref{eq23}) provides
us
with a definition of the Weyl product of two equivalent classes of Weyl
symbols.

From (\ref{eq23}), assuming that the Weyl symbols are smooth functions in
$\Re^{2n}$, we can derive a generalized Groenewold's formula for the
Weyl product in the form:

\begin{equation}
\left. (A*B)(z) \stackrel{\cal{M}}{=} \exp({i\hbar \over 2}
\sum_{i,j=1}^{2n} J^{D}_{ij}
\frac{\partial}{\partial \zeta_{i}}
\frac{\partial}{\partial \xi_{j}}) A(\zeta) B(\xi) \right|_{\zeta =\xi =z}
\label{eq24}
\end{equation}

In particular, it is not difficult to see that:

\begin{equation}
z_{i}*z_{j} = z_{i}z_{j} + {i \hbar \over 2} J^{D}_{ij}
\label{eqD}
\end{equation}

The Moyal bracket \cite{osborn} between two Weyl symbols is defined as
follows:

\begin{equation}
\{A, B\}_{M} = {1 \over i\hbar} [A*B - B*A]
\label{eq25}
\end{equation}

From (\ref{eq23}) and (\ref{eq25}) we obtain:

\begin{equation}
\{A, B\}_{M} \stackrel{\cal{M}}{=} (\pi \hbar)^{-2n} ({\hbar \over
2})^{2m-1} \int d\mu(\zeta) d\mu(\xi) A(\zeta) B(\xi) \sin({2 \over
\hbar} (\zeta - z) J^{D} (\xi - z))
\label{eq26}
\end{equation}

If we assume that the Weyl symbols are smooth functions in $\Re^{2n}$,
then, using the generalized Groenewold formula (\ref{eq24}) we can express
the
Moyal bracket between two symbols as follows:

\begin{equation}
\left.  \{A,B\}_{M} \stackrel{\cal{M}}{=} {2 \over \hbar} \sin({\hbar
\over
2} \sum_{i,j=1}^{2n} J^{D}_{ij} \frac{\partial}{\partial \zeta_{i}}
\frac{\partial}{\partial \xi_{j}}) A(\zeta) B(\xi) \right|_{\zeta =\xi =z}
\label{eq27}
\end{equation}

It is not difficult to prove that the new Moyal brackets satisfy the
Jacobi
identity.

If we assume that the Weyl symbols $A$ and $B$ are semiclassically
admissible \cite{osborn}, in other words, if we assume that $A$ and $B$
can be written as \cite{osborn}:

\begin{equation}
A(\zeta) = A_{c}(\zeta) + \sum_{r=1}^{\infty} \frac{\hbar^{r}}{r!}
a_{r}(\zeta),
\label{eq28}
\end{equation}

\begin{equation}
B(\zeta) = B_{c}(\zeta) + \sum_{r=1}^{\infty} \frac{\hbar^{r}}{r!}
b_{r}(\zeta),
\label{eq29}
\end{equation}

\noindent then, the new Moyal bracket (\ref{eq27}) gives:

\begin{equation}
\{A,B\}_{M} \stackrel{\cal{M}}{=} \sum_{i,j=1}^{2n} J^{D}_{ij} \nabla_{i}
A_{c} \nabla_{j} B_{c} +
O(\hbar^{2}) \stackrel{\cal{M}}{=} \{A_{c}, B_{c}\}_{D} + O(\hbar^{2})
\label{eq30}
\end{equation}

To zeroth order in Planck's constant, the new Moyal bracket is equal to
the Dirac bracket.

It is not difficult to see that:

\begin{equation}
\{z_{i}, z_{j}\}_{M} = J^{D}_{ij}
\label{eqE}
\end{equation}

The quantum evolution of the system is described by the equations:

\begin{equation}
\dot{z}_{i} = \{z_{i},H\}_{M}
\label{eq31}
\end{equation}

\noindent where $H$ is a Weyl symbol corresponding to the quantum operator
$\hat{H}_{c}$ in (\ref{eq17}).

Notice that, under the strong assumption of linearity of the constraints,
it is straightforward to give a Weyl-Wigner-Moyal formulation directly on
the reduced phase space $\cal{M}$. It is not difficult to prove that our
definitions (\ref{eq20}), (\ref{eq23}) and the generalized Groenewold
formula
(\ref{eq24}) are consistent with such a WWM formulation on the reduced
phase
space.

The problem of quantization on non-trivial reduced phase spaces has been
studied by several authors (see for example \cite{marinov,karasev}).

\vskip 1cm

It is my pleasure to thank M. Karasev, M. Marinov, F. Molzahn and T.
Osborn for very useful discussions and comments.


\vskip 1cm


\begin{thebibliography}{99}

\bibitem{weyl} H. Weyl, {\it The Theory of Groups and Quantum Mechanics}
(Dover Publications, New York, 1931).

\bibitem{wigner} E.P. Wigner, Phys. Rev. {\bf 40} (1932), 749.

\bibitem{moyal} J.E. Moyal, Proc. Cambr. Phil. Soc. {\bf 45} (1949), 99.

\bibitem{osborn} T.A. Osborn and F.H. Molzahn, Ann. Phys. {\bf 241}
(1995), 79.


\bibitem{little} R.G. Littlejohn, Phys. Rep. {\bf 138} (1986), 193.

\bibitem{muller} M. Muller, J. Phys. A{\bf 32} (1999), 1053.

\bibitem{antonsen} F. Antonsen, {\it Deformation Quantization of
Constrained Systems} (1997), gr-qc/9710021.

\bibitem{cabo} A. Cabo and D. Louis-Martinez, Phys. Rev. D{\bf 42}
(1990),2726.

\bibitem{lusanna} M. Chaichian, D. Louis-Martinez and L. Lusanna, Ann.
Phys. {\bf 232} (1994), 40.

\bibitem{gitman} D.M. Gitman and V.I. Tyutin, {\it Quantization of Fields 
with Constraints} (Springer-Verlag, Berlin, 1990).

\bibitem{dirac} P.A.M. Dirac, {\it Lectures in Quantum Mechanics} (Belfer 
Graduate School of Science, Yeshiva University, New York, 1964).


\bibitem{marinov} D. Bar-Moshe and M.S. Marinov, {\it Berezin
Quantization and Unitary Representations of Lie Groups} (1994),
hep-th/9407093.

\bibitem{karasev} M. Karasev, Contemporary Mathematics {\bf 179} (1994),
83.



\end{thebibliography}
\end{document}